\documentclass{Interspeech2024}
\usepackage{graphicx}
\usepackage{multirow}
\usepackage{float}
\usepackage{cite}

\usepackage[font=footnotesize,skip=0pt]{subcaption} 

\interspeechcameraready

\title{Towards Naturalistic Voice Conversion: NaturalVoices Dataset with an Automatic Processing Pipeline}

\name[affiliation={1}]{Ali N.}{Salman}
\name[affiliation={1}]{Zongyang}{Du}
\name[affiliation={1}]{Shreeram Suresh}{Chandra}
\name[affiliation={1}]{İsmail Rasim}{Ülgen}
\name[affiliation={1}]{Carlos}{Busso}
\name[affiliation={1}]{Berrak}{Sisman}

\address{
  $^1$Department of Electrical and Computer Engineering, The University of Texas at Dallas
  }
\email{ali.salman@utdallas.edu, zxd220002@utdallas.edu, sxc220013@utdallas.edu, ismailrasim.ulgen@utdallas.edu, busso@utdallas.edu, berrak.sisman@utdallas.edu}


\begin{document}

\maketitle

\begin{abstract}
Voice conversion (VC) research traditionally depends on scripted or acted speech, which lacks the natural spontaneity of real-life conversations. While natural speech data is limited for VC, our study focuses on filling in this gap. We introduce a novel data-sourcing pipeline that makes the release of a natural speech dataset for VC, named NaturalVoices. The pipeline extracts rich information in speech such as emotion and signal-to-noise ratio (SNR) from raw podcast data, utilizing recent deep learning methods and providing flexibility and ease of use. NaturalVoices marks a large-scale, spontaneous, expressive, and emotional speech dataset, comprising over 3,800 hours speech sourced from the original podcasts in the MSP-Podcast dataset. Objective and subjective evaluations demonstrate the effectiveness of using our pipeline for providing natural and expressive data for VC, suggesting the potential of NaturalVoices for broader speech generation tasks.

\end{abstract}
\section{Introduction}
\label{ssec:data_collect}

VC aims to convert one speaker's voice to sound like that of target speaker while preserving linguistic content \cite{qian2019autovc}. It has various applications such as movie dubbing, intelligent dialogue systems, real-time voice cloning, voice assistants, and conversational agents.

Most VC frameworks \cite{Kaneko_2028,qian2019autovc} are typically trained using scripted or acted corpus \cite{yamagishi2019cstr, zhou2022emotional}, resulting in the generation of high-quality speech predominantly with a reading or acting style. However, the process of collecting acted data for VC is labor-intensive and time-consuming, often burdened with inefficiencies. Real-life speech, in contrast, is spontaneous and encompasses various speaking styles \cite{du2021expressive}, emotional expressions\cite{zhou2022emotional}, nonverbal cues \cite{nguyen2023expresso} such as laughter, and lip smacks, as well as dysfluencies like repetitions, hesitations or interruptions \cite{10389771}. Therefore, the progress of VC models requires obtaining a more diverse dataset that reflects the richness, complexity, and expressiveness of spontaneous human speech. 

\begin{table*}
    \centering
    \caption{A comparison of the NaturalVoice dataset with other common VC datasets. The NaturalVoice dataset marks as the first large-scale, spontaneous, expressive, and emotional speech dataset for VC. }
    \vspace{-3mm}
    \fontsize{8}{9}\selectfont
    \scalebox{1}{\begin{tabular}{c|c|c|c|c|c|cc|c}
    \hline
    \multirow{2}{*}{} & \multirow{2}{*}{Speech Type} & \multirow{2}{*}{Total Hour} & \multirow{2}{*}{\# of Speakers} & \multirow{2}{*}{SNR} & \multirow{2}{*}{\begin{tabular}[c]{@{}c@{}}Sound Event \\ Categories\end{tabular}} & \multicolumn{2}{c|}{Emotion}                 & \multirow{2}{*}{Text} \\ \cline{7-8}
        & & & & & & \multicolumn{1}{c|}{Categories} & Attributes & \\ \hline
        VCTK \cite{Veaux_2017} & Read & 44 & 110 & No & No & \multicolumn{2}{c|}{ Neutral Speech}  & Transcript\\ 
        ESD \cite{zhou2022emotional}& Read & 15 & 10 & No & No & \multicolumn{1}{c|}{Yes} & No & Transcript\\ 
        \textbf{NaturalVoices}& Spontaneous & 3846 & $>$2467  & Yes & Yes& \multicolumn{1}{c|}{Yes}  & Yes & ASR \\ \hline
    \end{tabular}}

    \vspace{-3mm}
    \label{tab: datacomparision}
\end{table*}

In this paper, we introduce an automated method for sourcing data from podcasts for VC. Podcasts offer distinct advantages over other in-the-wild data sources such as YouTube data. Unlike the diverse and often noisy background of other data sources, podcasts generally have higher-quality audio recordings with clearer speech, making them particularly suitable for VC tasks. Additionally, podcasts provide a diverse range of speakers, ensuring that there's enough speech data for each speaker to effectively model their identity. This abundance of varied speakers and ample speech data for each speaker enables the successful transformation of speaker identity for VC task.


The components of our pipeline are depicted in Figure \ref{fig:pipelinestruct}. Leveraging state-of-the-art deep learning techniques from various speech tasks including diarization, automatic speech recognition, speaker recognition, and speech emotion recognition, our automated data sourcing pipeline presents an innovative solution to the challenges of gathering spontaneous and naturalistic data for VC. Applying this pipeline to the original podcasts in the MSP-Podcast dataset \cite{Lotfian_2019_3}, we release a subset optimized for VC, known as NaturalVoices. Below, we outline several key advantages of our approach and the NaturalVoices dataset:

\begin{figure}[t]
\centering
\includegraphics[,clip,width=0.37\textwidth]{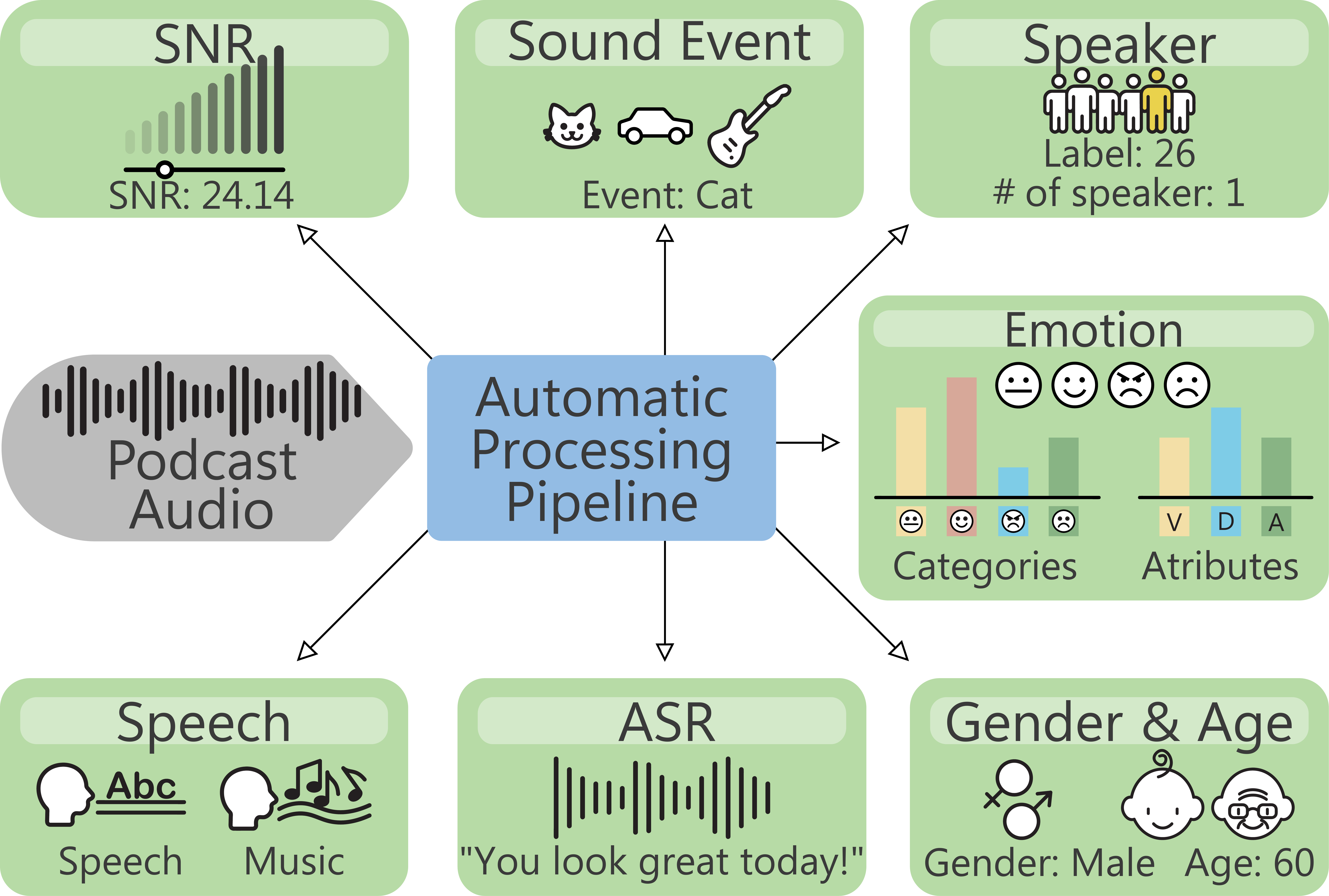}
\vspace{-2mm}
\caption{An illustration of our data sourcing pipeline with various modules.}
\vspace{-6mm}
  \label{fig:pipelinestruct}
\end{figure}

\noindent\textbf{Richness and Expressiveness:}
NaturalVoices comprises 3,846 hours of speech data from numerous speakers, capturing natural emotional expressions, diverse communication styles, nonverbal vocal cues, and a variety of background sounds in authentic recording conditions. Our pipeline provides transcripts,  speaker details (such as label and gender), signal-to-noise ratio (SNR), emotion attributes (arousal, dominance, valence) and emotion categories (neutral, angry, happy, sad), and sound event categories like laughter and animal sounds (e.g., rooster crowing). With such rich and expressive data provided by our pipeline, NaturalVoices can be utilized across a range of tasks in speech synthesis.

\noindent\textbf{Flexibility and Easy-to-Use :}  Our dataset and pipeline are publicly accessible\footnote{https://NaturalVoices.github.io/NaturalVoices/}. Utilizing our pipeline, users can easily filter NaturalVoices speech data based on specific criteria such as SNR, emotion states and gender. This simplifies dataset utilization across diverse applications.

The rest of this paper is structured as follows: Section~\ref{related_works} provides an overview of related work. In Section~\ref{pipeline}, we outline the details of our proposed automatic data-sourcing pipeline. Section~\ref{dataset} introduces the NaturalVoices dataset and explores its potential applications in other speech tasks. Experimental results are presented in Section~\ref{experiments}. Finally, Section~\ref{conclusion} summarizes our findings and concludes the study.

\section{Related Work}
\label{related_works}

\subsection{Emotional Speech Datasets}

\label{emotion_data}
Emotions play a vital role in human communication, yet understanding and synthesizing emotional speech remains challenging due to the nuanced and intricate expressions found in real-life conversations.   Speech emotion recognition (SER) is an important task for emotion understanding.  Existing emotional speech datasets for SER \cite{cao2014crema,adigwe2018emotional} often rely on scripted recordings, resulting in overemphasized emotions that differ from authentic emotional nuances found in natural conversation. Alternatively, collecting other SER datasets \cite{Busso_2008,Busso_2016} through conversational improvisations for spontaneous speech is labor-intensive and costly \cite{zhou2022emotional}. However, as highlighted by Zhou et al. \cite{zhou2022emotional}, these SER emotional datasets often lack lexical variability and may contain external noise and overlapping speech, making them unsuitable for VC.


Lotfian et al. \cite{Lotfian_2019_3} introduced the MSP-Podcast dataset for SER, consisting of retrieved speech segments with annotated emotion categories and attributes from podcast recordings. These podcasts contain natural conversations among diverse speakers discussing various topics. This diversity ensures large lexical and speaker variabilities, crucial for VC datasets. Inspired by this, we applied our pipeline to create NaturalVoices for VC task, without relying on annotations or labels from MSP-Podcast. 
\vspace{-2mm}
\subsection{Datasets for Voice Conversion}
VC frameworks require data from multiple speakers to model various aspects of speech, including speaker identity \cite{Kaneko_2028,qian2019autovc}, speaking style \cite{10243636}, and emotional information \cite{du2021expressive}.
VC datasets often consist of read speech, lacking natural spontaneity. For instance, the CSTR VCTK corpus \cite{Veaux_2017} contains recordings from 110 English speakers, and similarly, other VC datasets  \cite{Kominek2004TheCA,lorenzo2018voice,zhao2020voice,toda2016voice} primarily provide only neutral speech. The ESD dataset \cite{zhou2022emotional} includes emotional speech from 10 speakers with various emotional states, and has been used widely in VC. However, ESD is limited in size and consists solely of acted speech. To address the need for more diverse datasets reflecting spontaneous human speech, this paper introduces NaturalVoice, a large, natural, and emotional dataset for VC applications.

\section{Automatic Data Sourcing Pipeline}
\label{pipeline}


In this section, we introduce our automatic data-sourcing pipeline inspired by \cite{Upadhyay_2023_2}. Initially, our pipeline divides each podcast from the MSP-Podcast dataset into segments of appropriate duration. Subsequently, it automatically annotates these segments with transcripts, speaker details, SNR, emotion attributes, and sound event categories using different modules, as illustrated in Figure \ref{fig:pipelinestruct}. Our pipeline offers users the ability to easily and flexibly filter our dataset according to their preferences.

\noindent\textbf{Diarization and ASR Module:}
\label{pipeline_diarization}
We utilize the Faster Whisper model, an accelerated version of Whisper \cite{Radford_2022} combined with CTranslate2, to segment the audio of each episode into short segments (4-6 seconds) and provide accurate transcripts for each segment. However, minor misalignments persist in the model at segment start and end times. To resolve this, we adjust segment durations based on intervening silence: extending adjacent segments by 0.25 seconds each if silence exceeds 0.5 seconds, or merging segments for shorter silences. 
 We then use the Montreal Forced Aligner (MFA) \cite{Chodroff_2023} to align audio samples with transcriptions, generating phone-level alignments using word-to-phoneme mappings.

\noindent \textbf{Speech Detection Module: }
\label{pipeline_speech}
Some audio segments contain music instead of speech, which is usually undesired in many VC and speech synthesis applications, as the music can act as noise during training these models. We employ a temporal convolutional network  \cite{Lemaire_2019} to distinguish between speech and music.

\noindent \textbf{Speaker Recognition Module:}
\label{pipeline_asr}
A crucial aspect of training VC models is maintaining a unique and distinct speaker set to accurately represent speaker characteristics. We address this in two steps: Firstly, we utilize PyAnnote \cite{Bredin_2023, Plaquet_2023} to identify 'local speakers' within individual audio files. Secondly, we consolidate 'local speakers' from different files into 'global speakers' through manually annotating a single segment per file. This process creates a unique 'global speaker' set  for overlap-free applications and offers flexibility with 'local speakers'.

\noindent \textbf{Gender Classification and Age Module: }
\label{pipeline_gender_age}
We employ a model \cite{Felix_2023} for age and gender prediction, providing additional information for the speaker in each segment. 

\noindent \textbf{Emotion Attribute and Category Module: }
\label{pipeline_emotion}
We employ two high-performing emotion detection models. The first, PEFT-SER \cite{Feng_2023}, is a categorical model based on WavLM and LORA, trained to identify primary emotion classes within speech utterances, including neutral, happiness, sadness, and anger. The second is a regression-based WavLM model \cite{Goncalves_2024}, designed to assess the emotional spectrum of valence (ranging from negative to positive), arousal (from calm to excited), and dominance (from weak to strong) in each audio segment, thus providing a broader emotional context.

\noindent \textbf{SNR Module}:
\label{pipeline_snr}
SNR quantifies the level of desired signal relative to background noise in an audio recording. 
We estimate the SNR value for each audio segment using WADA-SNR \cite{Kim_2008_2}.

\noindent \textbf{Sound Event Detection Module}
\label{pipeline_audio_classification}
We use the AST model \cite{Gong_2021} to predict over 500 different sounds (e.g., honking, alarm, animal noises, etc) for each segment. By using this model we can filter the data to include or exclude any type of background or abnormal sounds.



\begin{figure}[!h]
    \centering
    \begin{subfigure}{0.4\textwidth}
        \includegraphics[width=\columnwidth]{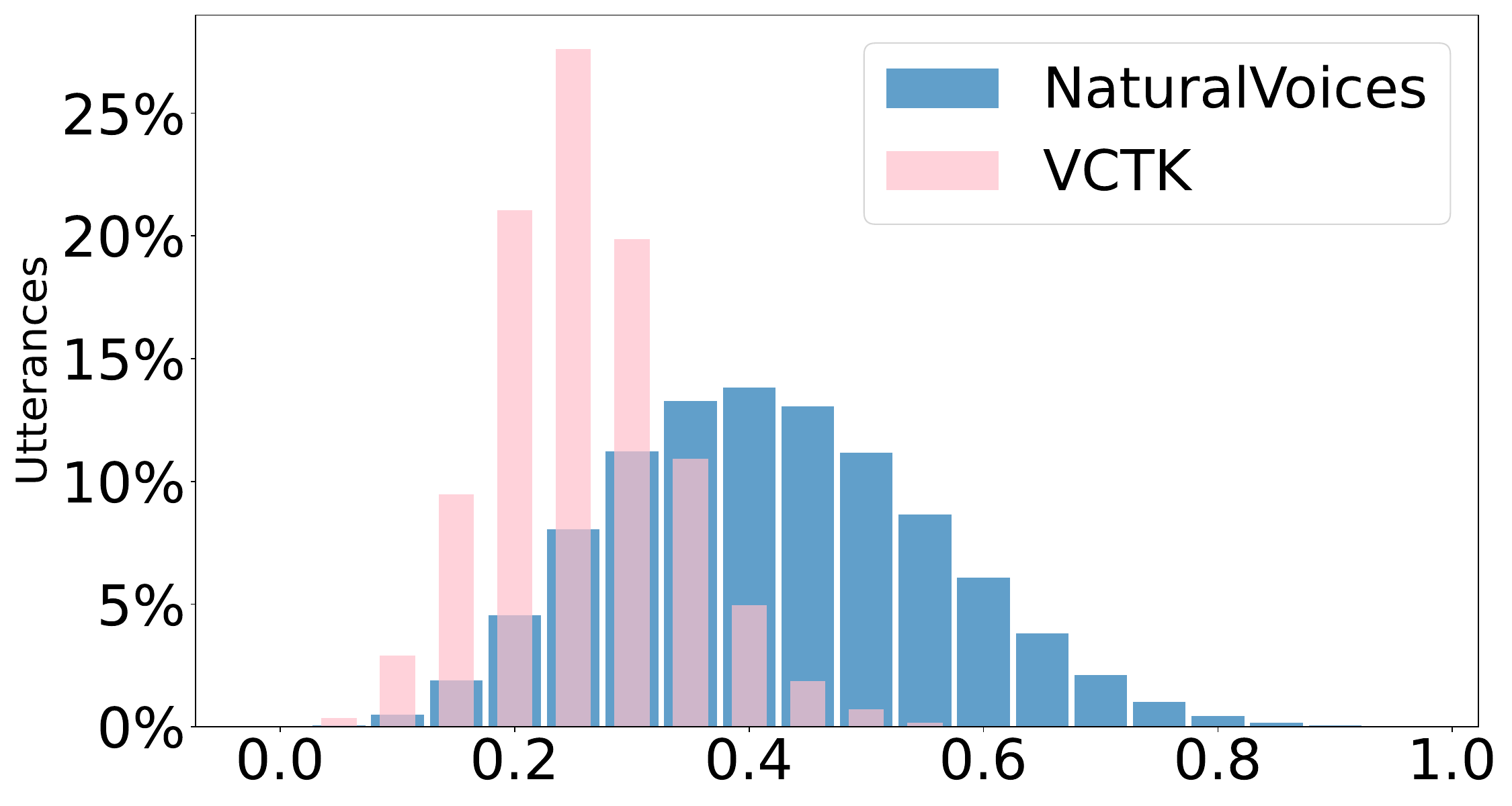}
        \caption{Arousal}
        \label{fig:arousal}
    \end{subfigure}
      \begin{subfigure}{0.4\textwidth}
        \includegraphics[width=\columnwidth]{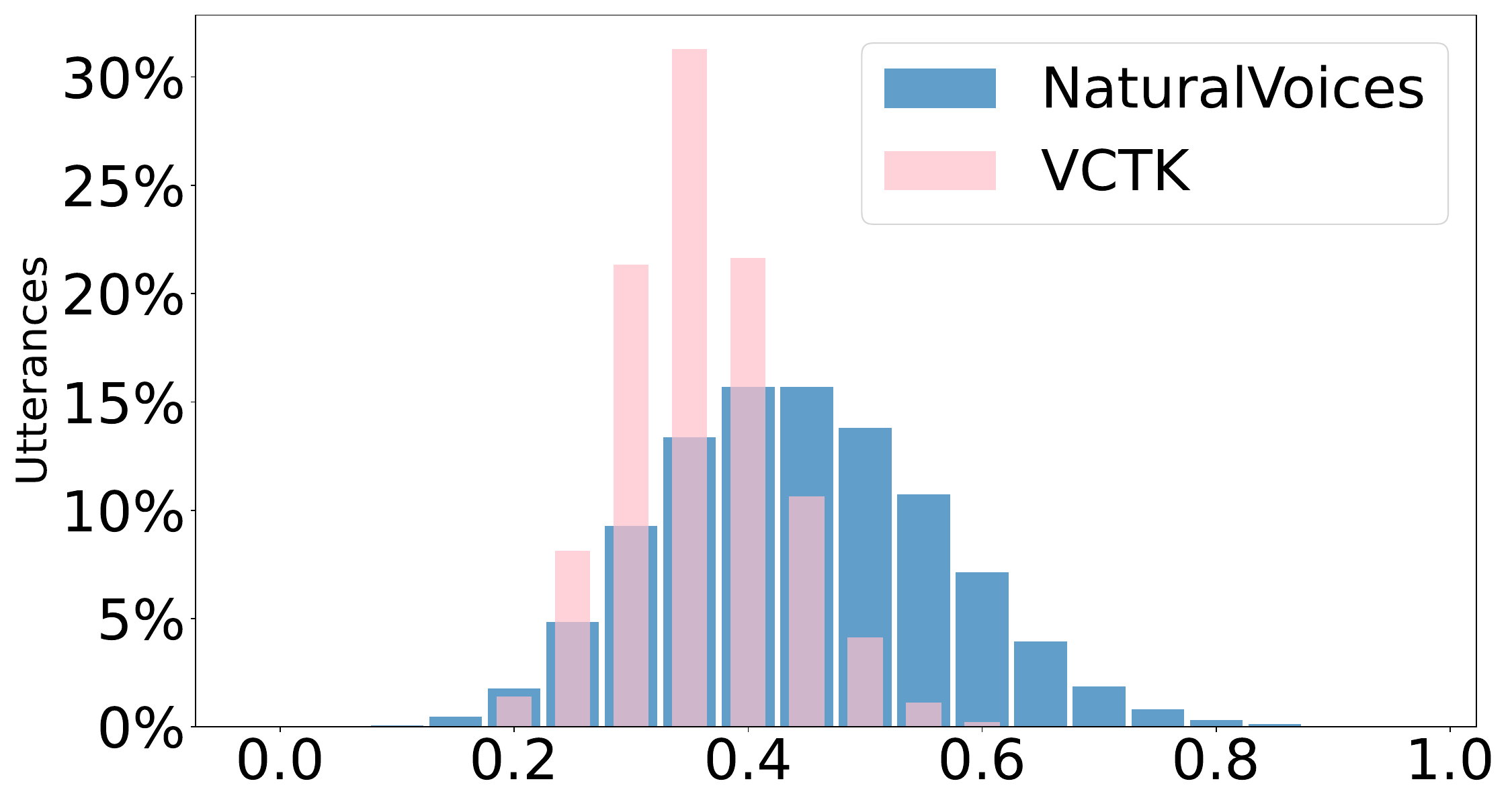}
        \caption{Dominance}
        \label{fig:dominance}
    \end{subfigure}
    \begin{subfigure}{0.4\textwidth}
        \includegraphics[width=\columnwidth]{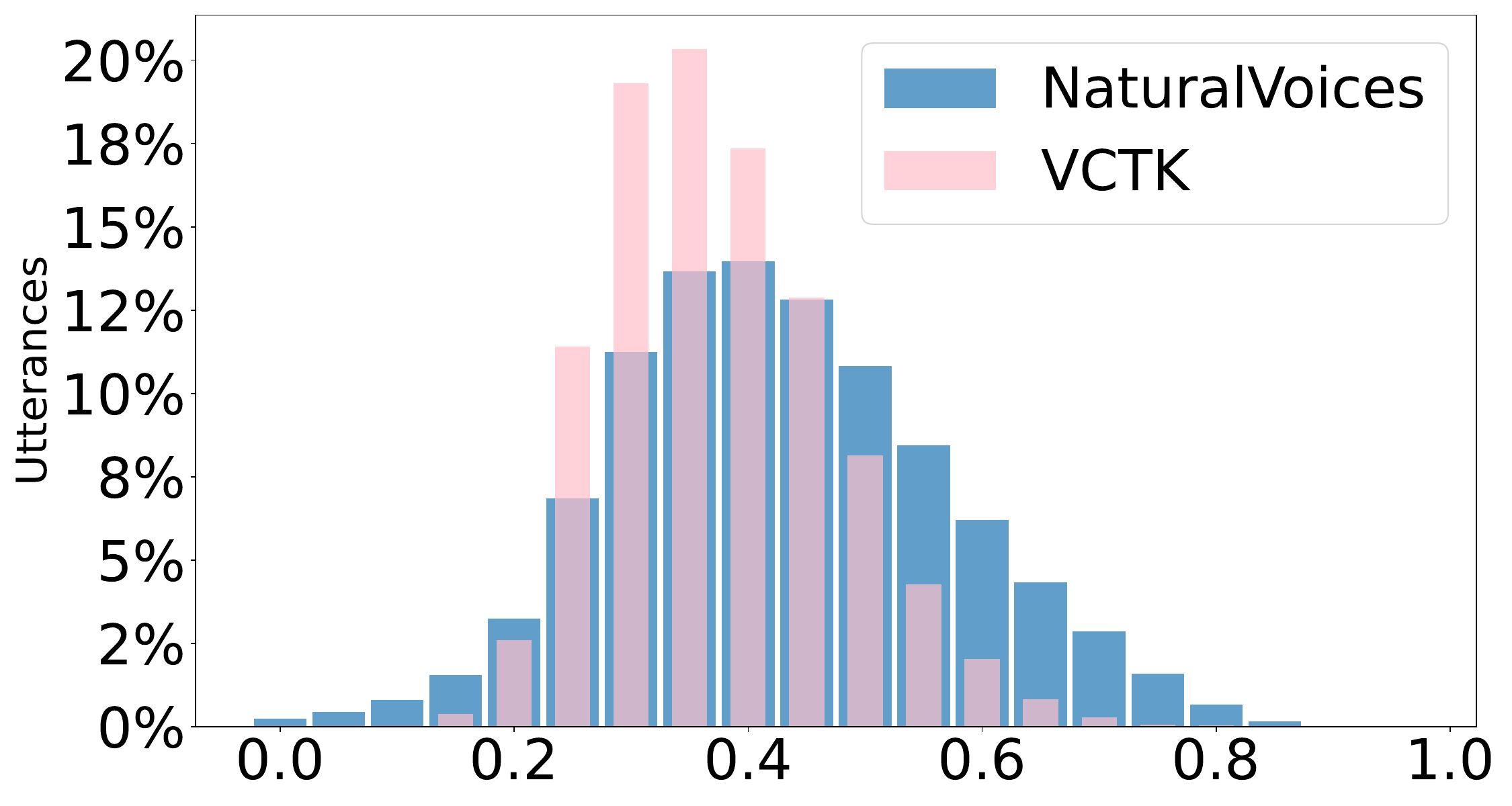}
        \caption{Valence}
        \label{fig:valence}
    \end{subfigure}
  \begin{subfigure}{0.4\textwidth}
        \includegraphics[width=\columnwidth]{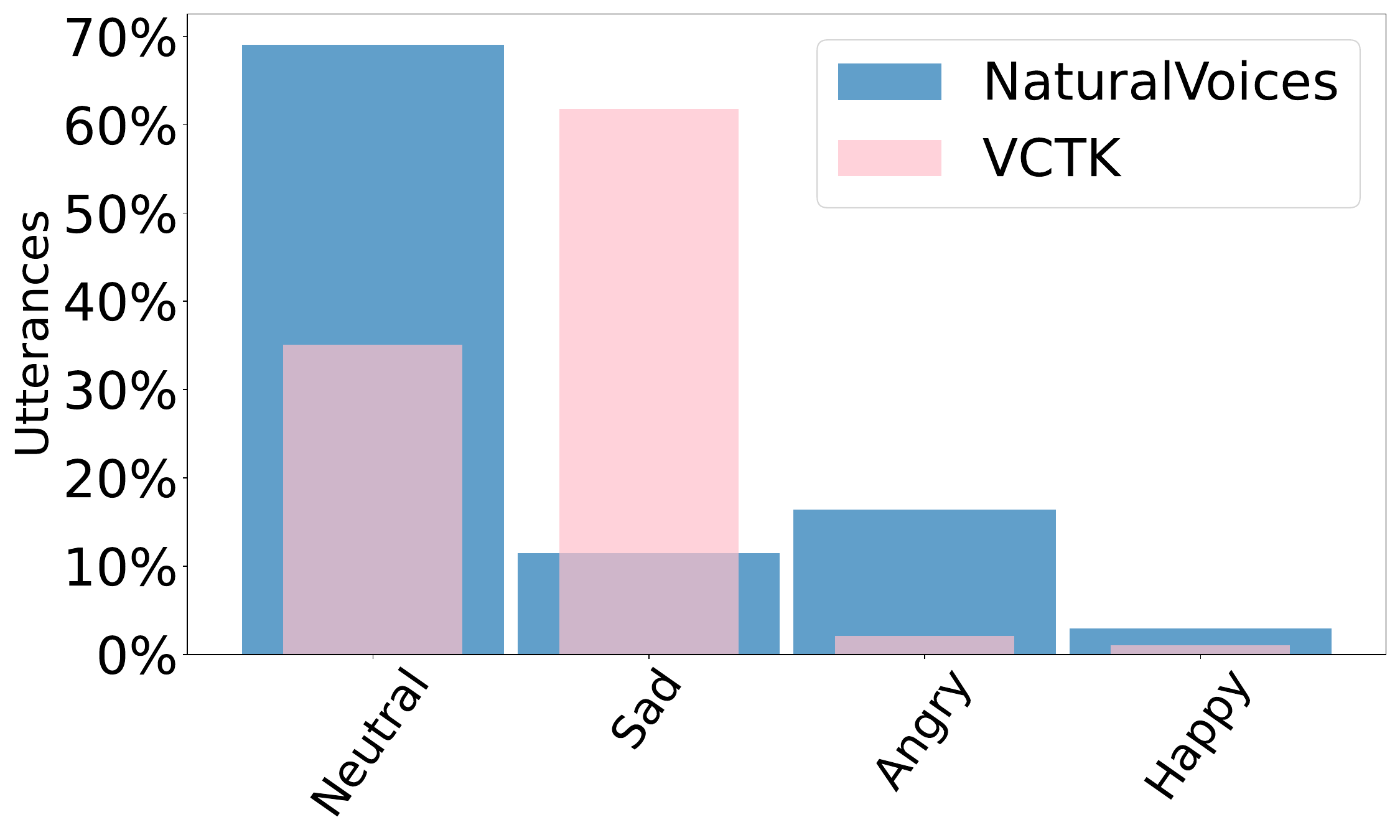}
        \caption{Categorical}
        \label{fig:categorical}
    \end{subfigure}
    \vspace{-2mm}
        \caption{Emotion attribute (arousal, dominance, valence) and category (neutral, sad, angry, happy) distribution in NaturalVoices versus VCTK.}
    \label{fig:emotion_distribution}
    \vspace{-7mm}
\end{figure}

\begin{figure}[t]
\centering
\includegraphics[width=0.40\textwidth]{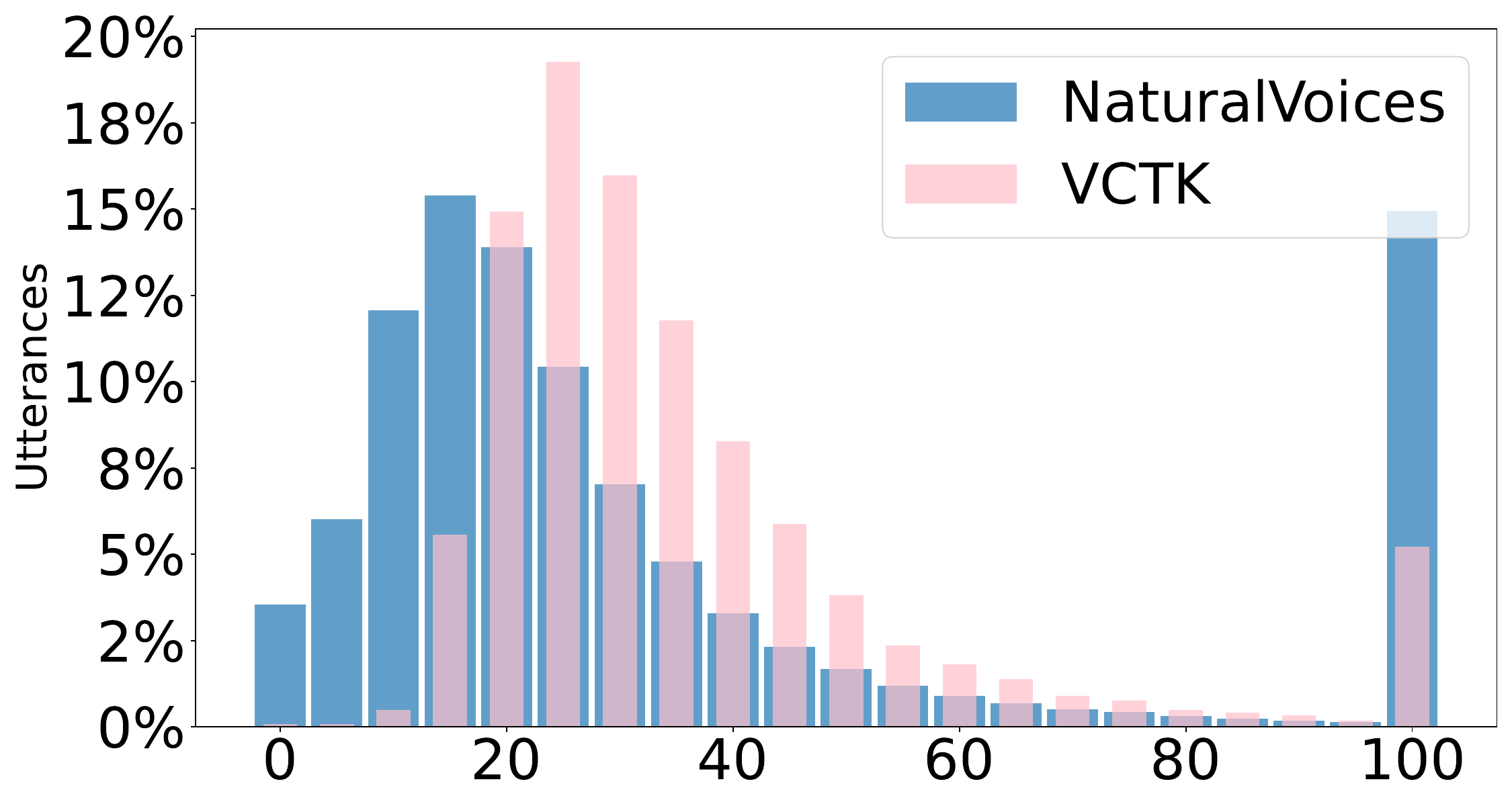}
\vspace{-3mm}
\caption{SNR distribution in NaturalVoices and VCTK.}
  \label{fig:snr}
\vspace{-7mm}
\end{figure}

\section{NaturalVoices Dataset}
\label{dataset}

Our research introduces a novel pipeline and the creation of publicly available, large-scale speech dataset, named NaturalVoices. NaturalVoices includes 3.8k hours of speech with over 2 million utterances, averaging 6.67 seconds each. Within this, 2.6k hours contain single-speaker speech, with 1.3k hours labeled for 2,467 speakers. The dataset includes 1,115 female speakers and 1,338 male speakers. All data is down-sampled to 16kHz. For each utterance, we provide other automatically annotated information obtained from the pipeline that are 1) Signal-to-Noise Ratio, 2) emotion attributes (arousal, dominance, valence), 3) emotion categories (neutral, angry, happy, sad), 4) speech or music classification, and 5) sound event categories. With this dataset, we aim to facilitate advancements in speech processing, in particular in expressive speech synthesis and VC. 
\vspace{-3mm}
 

\subsection{Analysis}


In our study, we conducted a comparative analysis between our dataset, NaturalVoices, and two widely used VC datasets: ESD \cite{zhou2022emotional} and VCTK \cite{Veaux_2017}, as summarized in Table 1. The results indicate that our dataset contains more natural and spontaneous speech, along with additional automatically annotated information. Compared to ESD, our dataset has a larger scale, including more speakers and greater lexical variability. Compared to VCTK, our dataset is more expressive, larger in scale, and includes more speakers. In the following sections, we provide a detailed comparison of VCTK from various perspectives.


Figures \ref{fig:arousal}, \ref{fig:dominance}, and \ref{fig:valence} provide visual representations of the distribution of emotion attributes – arousal, dominance, and valence – within both the VCTK and NaturalVoices datasets. Arousal reflects the degree of excitement or calmness, dominance indicates the intensity of an emotion, and valence denotes the positivity or negativity of an emotion \cite{Sridhar_2022}. As illustrated in Figures \ref{fig:arousal}, \ref{fig:dominance}, and \ref{fig:valence}, NaturalVoice exhibits a broad range across these three emotion attributes, indicating a higher level of expressiveness and the presence of stronger emotions. Conversely, VCTK demonstrates less variation, predominantly skewing towards negative attributes.

Moreover, categorical analysis highlights our dataset's wider spectrum of emotions in Figure \ref{fig:categorical}, including a higher prevalence of neutral, angry, and happy expressions, alongside a significantly larger dataset compared to VCTK.
The SNR distribution in Figure \ref{fig:snr} highlights the diversity of NaturalVoices in terms of background noise and varied environments, with segments below 40 SNR and at high SNR levels of 100. Compared to VCTK, our dataset contains more samples in the high SNR range (above 100) and a broader distribution across SNR levels, including more samples in the 0-20 SNR range. VCTK, on the other hand, predominantly focuses on the 20-40 dB range, with fewer samples in the extreme SNR values. This diversity underscores our dataset's wide range of recording scenarios, including instances with challenging background noise levels and those with optimal signal-to-noise ratios.
\vspace{-2mm}
\subsection{Possible Applications}
Our dataset presents a wealth of opportunities for diverse applications in speech processing and machine learning. In this section, we explore its potential impact on speech synthesis and other speech tasks.

\noindent \textbf{Expressive Speech Synthesis and VC:} Our dataset clearly facilitates speech synthesis and VC research. Furthermore, it can be used for expressive VC \cite{du2021expressive} and emotional VC \cite{zhou2022emotional}. While expressive VC focuses on converting speaker identity for emotional speakers, emotional VC specifically targets the conversion of emotional states. By providing a rich collection of natural speech data with diverse emotional attributes, our dataset enables VC models to capture and model nuanced emotional expressions. Additionally, with additional automatically annotated text, researchers can leverage this dataset to enhance the expressiveness and emotional fidelity of synthesized speech by developing text-to-speech (TTS) models. 

\noindent \textbf{Noisy-to-Noisy VC and Noise Robust VC:} Our dataset's diverse range of SNRs makes it ideal for exploring noisy-to-noisy VC, where the goal is to conduct identity conversion while preserving background sounds \cite{10256118}. Researchers can use this dataset to develop robust VC models that are capable of handling noisy input conditions, leading to improved performance in real-world scenarios with varying levels of background noise.

\noindent \textbf{Spontaneous Speech Modeling and Generation:} Understanding, modeling and generating spontaneous speech pose ongoing challenges. Our dataset encompasses a diverse range of conversational styles, pauses, and speech disfluencies, capturing the natural variability present in spontaneous speech. Researchers can leverage this dataset to train models for spontaneous speech generation, as well as for tasks such as spontaneous speech recognition and understanding.

\noindent \textbf{Weakly-Labeled Supervised Training:} Given that all annotations in our dataset are generated by deep learning models, the presence of weakly-labeled data offers a unique opportunity for weakly-labeled supervised training approaches. In such cases, labels may be incomplete or noisy, yet researchers can effectively leverage this data using techniques like semi-supervised learning and self-training. By doing so, they can improve model performance and generalization on various tasks, including ASR, speaker diarization, and SER.

\vspace{-2mm}

\section{Experiments for VC}
\label{experiments}
In this work, our primary focus to explore the use of NaturalVoices in VC under various settings. We demonstrate the effectiveness of NaturalVoices across a range of VC scenarios. 
\vspace{-3mm}


\subsection{Experimental Setup and Evaluation Metrics}
In this study, we employ TriAANVC \cite{park2023triaan} as our chosen VC model. Employing an encoder-decoder architecture, it effectively disentangles content and speaker features. Through the Triple Adaptive Attention Normalization (TAAN) block, the model extracts detailed and global speaker representations via adaptive normalization, ensuring preservation of source content with siamese loss and time masking. TriAANVC showcases state-of-the-art performance in non-parallel any-to-any VC tasks, as evidenced by evaluations on the VCTK dataset. 

We trained TriAANVC on our dataset, and compared the results with those obtained from the VCTK dataset. For training details, we utilized a batch size of 100, conducted training over 500 epochs, employed the Adam optimizer with a learning rate of $10^-5$, and set model parameters to H = 256, C = 512, and L = 6, with CPC utilized as features. And we used the ParallelWaveGAN \cite{yamamoto2020parallel} vocoder trained on our dataset to generate utterances. For the VCTK dataset, we employed the pre-trained model \footnote{https://github.com/winddori2002/TriAAN-VC}. We randomly selected six speakers as seen speakers and six speakers as unseen speakers from their demo page.

For objective evaluation, we focus on two key metrics: speaker similarity and intelligibility, which are both crucial aspects in assessing the effectiveness of VC systems. 
Speaker similarity is measured using the acceptance rate from a Speaker Verification (SV) model, based on cosine similarity between embedding vectors of target and converted speech. The threshold is determined using the equal error rate from our NaturalVoices dataset. Word Error Rate (WER) and Character Error Rate (CER) assess script discrepancies, with the script for converted utterances obtained using a pre-trained Whisper model \cite{Radford_2022}. For subjective evaluation, we conducted a Mean Opinion Score (MOS) \cite{du2021expressive} listening test to evaluate speech quality and intelligibility. A total of 12 subjects participated in all experiments.

\subsection{Experimental Comparisons and Results}
We utilized automatic annotations from our pipeline to filter training data from NaturalVoice, aiming for settings similar to VCTK: multiple speakers with neutral speech. This portion of our dataset was then employed to train TriAANVC. Since our dataset is out of distribution compared to VCTK, we also trained a vocoder with our dataset to enhance performance. Consequently, we obtained speech samples generated by two vocoders: one pre-trained on VCTK (denoted as NaturalVoices$_{VCTK}$), and the other trained on our dataset (referred to as NaturalVoices). The results presented in Table \ref{tab:obj} demonstrate that NaturalVoices achieve better performance in terms of speaker similarity and comparable performance for CER and WER, underscoring the efficacy of our dataset for VC tasks. Furthermore, training the vocoder on our dataset yields performance improvements, a noteworthy consideration given the dataset's out-of-distribution nature relative to VCTK. Subjective evaluation results in Table \ref{tab:sub} also affirm that our datasets are suitable for VC.

We examined the benefits of dataset size expansion by filtering a larger training dataset, denoted as NaturalVoice$_{-Large}$, to train TriAANVC. Table \ref{tab:obj} reports that a larger dataset enhances speaker identity conversion in the VC model. Given the spontaneous nature of NaturalVoices, increasing the size of the training data presents a challenge for the model designed for acted speech, particularly in modeling linguistic information for spontaneous speech.

Additionally, we explored the effect of different SNR levels on the VC model. We filtered our data into two settings: Low SNR (0-20 dB) and high SNR (80-100 dB). As demonstrated in Table \ref{tab:snr}, we can observe that the performance of the VC model is affected by background noise. Our dataset contributes to the development of robust VC models by providing valuable training data. 
\vspace{-3mm}

\begin{table}[!]
    \caption{Objective results for TriAANVC with VCTK and NaturalVoices in S2S and U2U settings. }
    \vspace{-3mm}
    \fontsize{8}{9}\selectfont
    \scalebox{0.9}{\begin{tabular}{l@{\hspace{0.1cm}} |c@{\hspace{0.1cm}}|c@{\hspace{0.1cm}}|c@{\hspace{0.1cm}}|c@{\hspace{0.1cm}}|c@{\hspace{0.1cm}}|c}
    \hline
        \multirow{2}{*}{Training Data} & \multicolumn{2}{c|}{SV(\%)$\uparrow$} & \multicolumn{2}{c|}{CER(\%)$\downarrow$} & \multicolumn{2}{c}{WER(\%)$\downarrow$} \\ \cline{2-7} 
                                    & S2S  & U2U   & S2S  & U2U  & S2S  & U2U  \\ \hline
        VCTK\cite{Veaux_2017}        & 71.10     &80.30       &16.42      & 12.38    &25.74      & 19.82     \\ \hline
        NaturalVoices      & 93.65     &  89.16     &17.01      &18.55      &27.20      &30.43       \\
        NaturalVoices$_{vctk}$               &80.75      & 82.76      &19.31      &19.26      &30.70      &31.02      \\
        NaturalVoices$_{-Large}$    &96.78     &73.80       &19.68     &22.26      &30.37      &33.31      \\ \hline
    \end{tabular}}

\label{tab:obj}
\end{table}

\begin{table}[!]
    \centering
    \caption{Objective results for TriAANVC across varying SNR levels in a S2S setting.}
    \vspace{-3mm}
    \fontsize{8}{9}\selectfont
    \begin{tabular}{c|ccc}
    \hline
        & SV$(\%)\uparrow$ & CER(\%)$\downarrow$ & WER(\%)$\downarrow$ \\ \hline
        Low SNR    &  85.11  &   32.69  & 46.91    \\
        High SNR   & 93.65    &    17.01        & 27.20           \\ \hline
    \end{tabular}
    \label{tab:snr}
\end{table}

\begin{table}[!]
\centering
\caption{MOS results for NaturalVoices with 95\% confidence interval.}
\vspace{-3mm}
\fontsize{8}{9}\selectfont
\begin{tabular}{c|cc}
\hline
             & Quality$\uparrow$ & Intelligibility$\uparrow$ \\ \hline
Generated Speech       &3.17$\pm$0.23         & 3.77$\pm$0.36                \\
NaturalVoices & 4.38$\pm$0.16        &   4.79$\pm$0.18             \\ \hline
\end{tabular}
\vspace{-4mm}
\label{tab:sub}
\end{table}

\section{Conclusion}
\label{conclusion}
In our paper, we introduce NaturalVoices, a novel large-scale spontaneous speech dataset comprising over 3,800 hours of diverse and emotional speech sourced from podcast data, leveraging an innovative data-sourcing pipeline. The pipeline operates concurrently to predict multiple labels relevant to speech synthesis and is designed to accommodate expansion as new and improved models emerge. 
Experiment results show VC model can generate natural and intelligible speech by using NaturalVoices, indicating its potential for broader speech generation applications. We will explore other expressive speech synthesis tasks with NaturalVoices in future work.



\newpage
\footnotesize
\bibliographystyle{IEEEtran}
\bibliography{refs,reference}

\end{document}